\documentclass[aip, amsmath, amssymb, reprint, floatfix]{revtex4-2}

\usepackage{graphicx}
\usepackage{bm}
\usepackage{etoolbox}

\usepackage[utf8]{inputenc}
\usepackage[T1]{fontenc}
\usepackage{mathptmx}
\usepackage{siunitx}
\usepackage{xfrac}
\usepackage{physics}
\usepackage{cleveref}
\usepackage{microtype}
\usepackage{soul}
\soulregister\cite7

\makeatletter
\def\@email#1#2{%
 \endgroup
 \patchcmd{\titleblock@produce}
  {\frontmatter@RRAPformat}
  {\frontmatter@RRAPformat{\produce@RRAP{*#1\href{mailto:#2}{#2}}}\frontmatter@RRAPformat}
  {}{}
}%
\makeatother
\begin{document}

\preprint{AIP/123-QED}

\title{HISOL: high-energy soliton dynamics enable ultrafast far-ultraviolet laser sources}
\author{Christian Brahms}
\email{c.brahms@hw.ac.uk}
\author{John C. Travers}
\affiliation{ 
    School of Engineering and Physical Sciences, Heriot-Watt University,\\Edinburgh, EH14 4AS, UK
}

\date{\today}

\begin{abstract}
    Ultrafast laser sources in the far ultraviolet (\SIrange{100}{300}{\nm}) have been the subject of intense experimental efforts for several decades, driven primarily by the requirements of advanced experiments in ultrafast science. Resonant dispersive wave emission from high-energy laser pulses undergoing soliton self-compression in a gas-filled hollow capillary fibre promises to meet several of these requirements for the first time, most importantly by combining wide-ranging wavelength tuneability with the generation of extremely short pulses. In this Perspective, we give an overview of this approach to ultrafast far-ultraviolet sources, including its historical origin and underlying physical mechanism, the state of the art and current challenges, and our view of potential applications both within and beyond ultrafast science.
\end{abstract}

\maketitle

\section{Introduction}
Ever since the invention of the laser itself, new laser sources have enabled new applications. One of the most important degrees of freedom in any laser application is the laser wavelength: it sets the spatial resolution of optical systems through the diffraction limit and, more importantly, it determines the fundamental nature of light-matter interaction. In ultrafast science, for instance, fundamental microscopic properties and dynamics of a wide variety of systems from single atoms to condensed matter are studied through the absorption of light and subsequent processes~\cite{maiuri_ultrafast_2020,chergui_ultrafast_2019,odate_brighter_2022,nisoli_attosecond_2017,geneaux_transient_2019}, both of which strongly depend on wavelength. The same variety of primary processes also underlies the need for laser sources at different wavelengths to process and modify industrial materials~\cite{hodgson_ultrafast_2021}.

Enabled primarily by continual progress in laser frequency conversion through nonlinear optics, current technology can deliver laser radiation in femtosecond and even attosecond pulses across a large range of the electromagnetic spectrum, from the terahertz~\cite{koch_terahertz_2023} to the x-ray~\cite{chini_generation_2014,schoenlein_recent_2019}. However, one part of the parameter space has stubbornly resisted the development of laser sources which fulfil all the requirements of cutting-edge research and technology despite significant efforts: ultrafast pulses in the far ultraviolet (UV) between around \SI{100}{\nm} and \SI{300}{\nm}. The far UV is of great interest for one important reason: virtually all matter resonantly absorbs radiation somewhere in this wavelength region, but unlike extreme-ultraviolet light at even shorter wavelengths, it does not necessarily cause direct single-photon ionisation. Because of this combination, far UV light holds huge potential for a wide range of applications from fundamental science to technology and biomedicine. In particular, much of the recent research at the cutting edge of ultrafast far-UV laser sources has been driven by the needs of time-resolved spectroscopy and imaging of dynamics in molecules~\cite{chergui_ultrafast_2019,nisoli_attosecond_2017,lepine_attosecond_2014,maiuri_ultrafast_2020,calegari_open_2023,odate_brighter_2022}. However, the same properties which make far-UV light useful also create profound challenges for the development of laser sources: most materials simply absorb any light which is generated, and even in those which do not, the proximity of electronic resonances leads to strong material dispersion and multi-photon absorption.

One approach to overcoming the far-UV bottleneck in laser technology is to exploit resonant dispersive wave (RDW) emission from higher-order solitons in gas-filled hollow capillary fibres (HCFs), a technique we call HISOL (for \textbf{Hi}gh-energy \textbf{Sol}itons). HISOL combines the high damage threshold and far-UV transparency of gas media, long interaction lengths enabled by waveguides, guidance of high-energy laser pulses in large-core HCFs, and the unique nonlinear evolution of ultrafast laser pulses in the higher-order-soliton regime. This combination has allowed us to efficiently and flexibly convert infrared laser pulses to wavelength-tuneable far-UV pulses with few-femtosecond duration and near-perfect beam properties. In this Perspective, we will describe the fundamental mechanism and historical origin of this frequency-conversion technique, review the state of the art and current challenges, and discuss our view of the future prospects and possible applications for far-UV RDW pulses. In reviewing previous work, we will focus on the most important milestones on the path towards the current state of the art; a recent comprehensive review of research on solitons in hollow-core waveguides of all types can be found in ref.~\cite{travers_optical_2024}.

\section{Ultrafast far-ultraviolet laser sources}
Cutting-edge ultrafast science requires an ultrafast far-UV laser source which combines wavelength tuneability over a large range with extremely short pulse duration down to the single- or few-femtosecond regime. Of course, for such a source to be useful in practice, all the usual requirements on ultrafast laser sources also apply: we would like short-term and long-term stability, compactness, temporal and spatial coherence, high beam quality, long component lifetime, power and energy efficiency, and average-power or repetition-rate scalability. A large variety of approaches to meeting these requirements have been investigated. Here we will briefly review some of the most successful ones, focusing in particular on efforts to reduce the duration of far-UV laser pulses.

As with other wavelength ranges, there are fundamentally two routes to accessing the far UV: direct emission from a far-UV laser or nonlinear frequency conversion. Very energetic far-UV laser pulses can be generated at free-electron laser facilities~\cite{ayvazyan_generation_2002,chang_tunable_2018}, but these sources have not reached very short pulse durations in this wavelength range, and their scale and cost makes access to them rare. At the tabletop scale, only two types of laser have directly generated ultrafast far-UV light to our knowledge. The output from ultrafast KrF excimer lasers~\cite{szatmari_high-brightness_1994} operating at \SI{248}{\nm} has been compressed to below \SI{20}{\fs} with nonlinear pulse compression in a gas-filled hollow capillary fibre~\cite{klein-wiele_hollow-fiber_2006-1}. Recently, a Ce:LiCAF laser oscillator at \SI{290}{\nm} with sub-\SI{100}{\fs} pulse duration was also demonstrated~\cite{sharp_generation_2021}. However, neither of these approaches have been pushed into the vacuum UV (below \SI{200}{\nm}) or to shorter pulse durations. Nonlinear frequency conversion from longer-wavelength lasers using a great variety of methods is a far more popular approach to reaching the far UV.

The most common method of generating far-UV laser pulses is to use nonlinear crystals, such as $\beta$-barium borate (BBO). Starting with a primary laser in the infrared, multiple stages of frequency up-conversion via sum-frequency generation and second-harmonic generation can yield far-UV laser pulses~\cite{ringling_tunable_1993-1,rotermund_generation_1998}. Combining this with optical parametric amplification (OPA) of a white-light seed~\cite{cerullo_ultrafast_2003} adds wide-ranging wavelength tuneability. However, the multi-stage up-conversion process results in low conversion efficiency, and the phase-matching bandwidth of nonlinear crystals limits the pulse duration in this approach to the range of tens of femtoseconds. Much shorter pulses in the visible spectral region can be created in non-collinear OPAs~\cite{wilhelm_sub-20-fs_1997} and these can then be frequency-doubled to sub-20-fs pulses between \SI{250}{\nm} and \SI{310}{\nm} in BBO~\cite{beutler_generation_2009}. The phase-matching bandwidth of BBO can be further extended through angular dispersion, which enables the generation of sub-10-fs pulses between \SI{270}{\nm} and \SI{340}{\nm}~\cite{baum_tunable_2004}.

Any nonlinear frequency conversion scheme is limited by the transparency of the nonlinear medium. In the case of BBO, this means that the shortest wavelength which can be generated is around \SI{190}{\nm}. Because the dispersion increases strongly as the wavelength approaches this limit, the phase-matching bandwidth is decreased, so the generation of short pulses is very challenging even before absorption becomes an issue. To push to even shorter pulses and shorter wavelengths, we have to abandon solid-state nonlinear media in favour of gases. Four-wave mixing (FWM) driven by two-colour pulses in gas-filled hollow-core waveguides~\cite{durfee_ultrabroadband_1997} can generate sub-10-fs pulses~\cite{durfee_intense_1999,kida_sub-10_2010,kida_generation_2011} and, through cascaded up-conversion, wavelengths as short as \SI{160}{\nm}~\cite{misoguti_generation_2001}. The cascade can be pushed much further, to \SI{69}{\nm}, but this has only been demonstrated with long pulses so far~\cite{couch_ultrafast_2020}. The FWM approach can be very efficient~\cite{misoguti_generation_2001,belli_highly_2019,belli_broadband_2020}, but the FWM process fixes the far-UV wavelength(s) based on those of the driving pulses. Additional nonlinear interactions (self- and cross-phase modulation) during the frequency conversion create a degree of tuneability~\cite{belli_broadband_2020}, but wide ranges can only be accessed by generating at least one of the two driving pulses with an OPA~\cite{jailaubekov_tunable_2005}. The same fundamental interaction can also be exploited in free space~\cite{fuji_generation_2007,beutler_generation_2010,ghotbi_generation_2010}. This generally produces somewhat longer pulses with low efficiency, but has the advantage that it can be employed directly inside a vacuum chamber, greatly reducing the amount of dispersion experienced by the far-UV pulse after generation. An alternative approach is to use the gas not as the frequency-conversion medium but instead to compress far-UV laser pulses to shorter durations. Pulse compression (at a fixed wavelength or with minor tuneability) through either self-phase-modulation~\cite{nagy_generation_2009,klein-wiele_hollow-fiber_2006-1} or cross-phase modulation from an infrared pulse~\cite{jiang_ultraviolet_2024} has yielded very short and high-energy far-UV laser pulses.

Perhaps the simplest gas-based frequency conversion scheme is also the one which has yielded the shortest pulses: direct third-harmonic generation in a gas target~\cite{reiter_route_2010,reiter_generation_2010,graf_intense_2008}. When driven by few-cycle infrared pulses, far-UV pulses with a duration below \SI{2}{\fs} can be generated in this way~\cite{galli_generation_2019}---the shortest far-UV laser pulses reported to date. However, because the interaction length is limited by the gas target, the conversion efficiency is low. Additionally, the far-UV wavelength can only be changed by tuning the driving laser, which is a formidable technical challenge, especially because few-cycle driving pulses are required. 

The key constraint shared by all sources we have discussed so far is that the wavelength and duration of the far-UV pulse are \emph{directly tied to those of the initial infrared driving pulses}: short far-UV pulses require short infrared driving pulses, and their wavelength is largely determined by the driving wavelength(s). It is by \emph{breaking these direct relationships} that RDW emission can surpass the performance of previous ultrafast far-UV sources, especially in terms of wavelength tuneability and pulse duration.

\section{Resonant dispersive wave emission}
\begin{figure}
    \includegraphics[width=8.5cm]{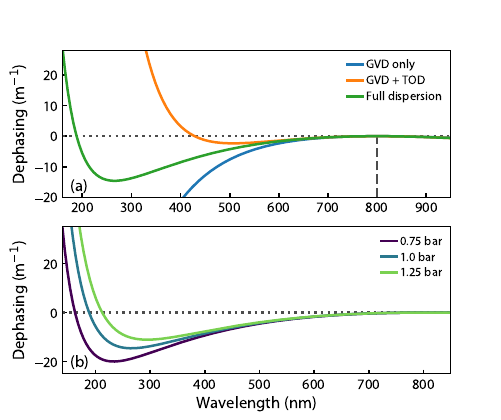}
    \caption{(a) Dephasing rate between a non-dispersing soliton at \SI{800}{\nm} and a dispersive wave in an HCF with \SI{125}{\um} core radius filled with helium at \SI{1}{\bar} pressure, shown at three levels of approximation: with second-order dispersion (GVD) only; with third-order dispersion (TOD) added; and with the full expression for the dispersion. The phase-matching point for RDW emission occurs where the dephasing vanishes. (b) Dephasing from the full expression for the dispersion for three different gas pressures in the same system, showing the wavelength tuneability of RDW emission.}
    \label{fig:phasematching}
\end{figure}

The key dynamical process underlying RDW emission in the UV is soliton self-compression. Solitons are a ubiquitous feature of nonlinear systems across all areas of physics and can result from the combination of many different fundamental phenomena. Optical temporal solitons specifically are formed by the interplay between the third-order nonlinearity (the optical Kerr effect) and the group-velocity dispersion when both are acting on a pulse travelling through a medium. If the group-velocity dispersion is anomalous (negative), it down-chirps the pulse: lower frequencies arrive later than higher frequencies. This counteracts the effect of the intensity-dependent refractive index, which creates an up-chirp instead. In a \emph{fundamental soliton}, nonlinearity and dispersion precisely balance each other, so that neither of the two effects cause their usual changes to the pulse: the nonlinearity does not lead to spectral broadening, and the dispersion does not stretch the pulse in time. The only change in the pulse is a simple uniform phase evolution. More interesting dynamics occur if the nonlinearity is stronger than the dispersion (which is most easily achieved by increasing the peak power). In such a \emph{higher-order soliton}, nonlinear spectral broadening does occur, and the resulting chirp is continually compensated by the dispersion to some degree. This reduces the pulse duration. Because this results in higher peak power, the effect intensifies in a positive feedback loop, leading to extreme self-compression of the pulse as it propagates. This is the time-domain analogue to self-focusing of laser beams in bulk media~\cite{zakharov_exact_1972, hasegawa_transmission_1973}. It is important to note that, although exact soliton propagation only occurs for a specific pulse shape (a hyperbolic secant envelope), most soliton effects can be readily observed by driving with different pulses~\cite{chen_nonlinear_2002}. Solitons can even emerge from noise-like input fields~\cite{gouveia-neto_soliton_1989}. This means that there is no fundamental difference between the driving pulses used in studies of soliton effects and those in other ultrafast experiments. \emph{Any} ultrafast laser pulse experiencing anomalous dispersion and positive third-order nonlinearity can be viewed as a combination of one or more overlapping fundamental optical solitons at the central frequency of the pulse and a weak non-solitonic component~\cite{zakharov_exact_1972, Mitschke2017}. Any such collection of fundamental solitons undergoes higher-order soliton evolution. The key question is whether the experimental parameters (e.g., medium length, losses, soliton order) allow for the observation of soliton \emph{effects}, such as self-compression and RDW emission.

Optical solitons in fibres were first proposed just over fifty years ago~\cite{hasegawa_transmission_1973} and demonstrated experimentally a few years later~\cite{mollenauer_experimental_1980}. RDW emission was first described soon after as a perturbation to the evolution of optical solitons, which were being considered as a potential information carrier in telecommunications~\cite{wai_nonlinear_1986}. When propagating in the vicinity of the zero-dispersion wavelength of an optical fibre, where higher-order dispersion effects are important, soliton pulses were found to break up, and this coincided with the appearance of a spectral sideband at higher frequencies. The mechanism by which energy can be transferred from the soliton to the dispersive wave was then studied in detail from a theoretical perspective~\cite{karpman_radiation_1993,akhmediev_cherenkov_1995, elgin_perturbative_1995}, but simple models have so far failed to capture the complex and subtle nonlinear dynamics. For example, there is as yet no reliable way of predicting the conversion efficiency of RDW emission without numerical simulations.

\begin{figure*}
    \centering
    \includegraphics[width=17cm]{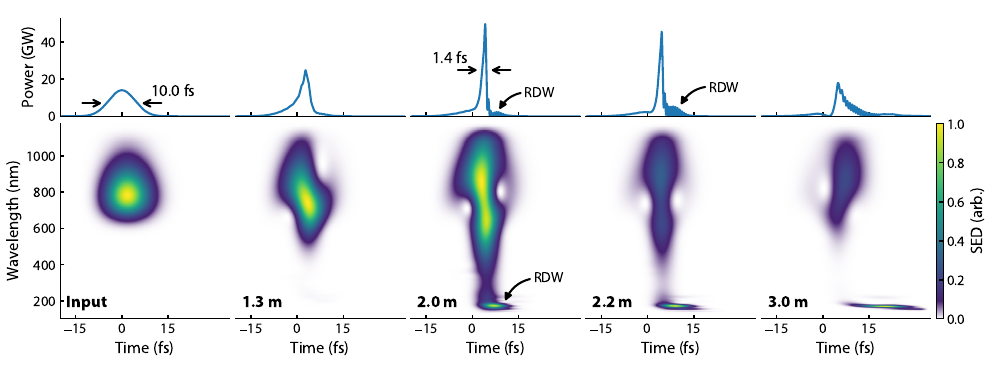}
    \caption{Simulated pulse profiles and spectrograms of a laser pulse undergoing soliton self-compression and RDW emission. The initial laser pulse is centred at \SI{800}{\nm} and has a duration of \SI{10}{\fs} and an energy of \SI{150}{\uJ}. The waveguide in this simulation is a 3-m-long HCF with \SI{125}{\um} core radius filled with helium at \SI{1}{\bar} pressure. (These are similar parameters to the first experimental demonstration of the HISOL concept~\cite{travers_high-energy_2019}.) Each spectrogram, computed with a Gaussian gate function (\SI{7.5}{\fs} full width at half maximum), shows the time-frequency distribution of the pulse at one position along the pulse propagation.}
    \label{fig:spectrograms}
\end{figure*}

Perhaps the simplest picture of the process can be drawn by considering a greatly simplified model of a higher-order soliton. Although RDW emission critically depends on the presence of higher-order dispersion, we first pretend that the self-compressing pulse effectively experiences no dispersion at all and evolves simply with a uniform nonlinear phase shift like a soliton. We then calculate the dephasing rate between this non-dispersing pulse and a general linear (that is, low-power) wave propagating in the same fibre. The result is shown in \cref{fig:phasematching}(a) for the cases of pure second-order dispersion, with third-order dispersion added, and with the full expression for the dispersion. The third-order term causes the propagation constants of the soliton and dispersive wave to cross; at this frequency, the two hypothetical pulses are phase-matched. \emph{If} the soliton bandwidth extends to this frequency, energy can be efficiently transferred to the phase-matched spectral band. Here, the self-compression process becomes important: RDW emission can be efficient even if the phase-matched frequency lies far outside the bandwidth of the initial driving pulse, because energy can reach the dispersive wave by spectral broadening. This process is visualised in the time-frequency domain in \cref{fig:spectrograms}: the initial infrared pulse undergoes extreme spectral broadening during soliton self-compression to sub-cycle duration, with the short-wavelength side enhanced by self-steepening. As a result, the spectrum reaches the phase-matched wavelength for RDW emission (in this case around \SI{170}{\nm}). After it is generated, the dispersive wave ``detaches'' from the main pulse due to a difference in group velocity and stretches in time due to dispersion.

We should note that, while it is instructive, the cartoonishly simple picture of the soliton evolution we have drawn above is not \emph{required} to derive the phase-matching condition. A more detailed analysis considering the cascaded four-wave mixing interactions which underlie spectral broadening via self-phase modulation comes to the same conclusion without considering soliton propagation at all~\cite{erkintalo_cascaded_2012}: when considering discrete frequency components, spectral broadening effectively mimics a higher-order nonlinearity, and the phase-matching condition for this hypothetical process is exactly the same as the one derived from our toy model here.

Shortly after the advent of photonic crystal fibre~\cite{knight_all-silica_1996}, which enabled soliton-based supercontinuum generation in the visible spectral region~\cite{ranka_visible_2000-1}, RDW emission was identified as the key mechanism by which the continuum was extended to short wavelengths~\cite{Husakou2001}. Over the next decade, the short-wavelength edge of supercontinuum generation in solid-core fibre was pushed ever further by engineering the dispersion profile of the waveguides~\cite{travers_blue_2010}, but it is ultimately limited by the absorption and strong dispersion of silica glass in the UV. As with other frequency-conversion schemes, generating very short pulses at very short wavelengths requires a gas medium. Although soliton self-compression in a gas was first observed in a hollow-core photonic bandgap fibre~\cite{ouzounov_generation_2003}, the guidance bandwidth was not sufficient to observe a dispersive wave in the UV. Antiresonant hollow-core fibres, on the other hand, exhibit ultrabroadband guidance from the UV to the infrared~\cite{benabid_stimulated_2002}. Far-UV RDW emission was first proposed in numerical simulations~\cite{im_high-power_2010} and demonstrated experimentally shortly after~\cite{joly_bright_2011}. These works already showed the two features which make for a unique far-UV light source: extremely short pulse duration and wide-ranging wavelength tuneability.

RDW emission in gas-filled hollow-core waveguides can efficiently generate very short, wavelength-tuneable far-UV pulses for two main reasons: firstly, the broad transparency window allows for extreme self-compression to multi-octave, sub-cycle-duration pulses \emph{before} RDW emission occurs, as shown in \cref{fig:spectrograms}. This means that the frequency conversion process is tightly confined in time and necessarily creates a short pulse, at least initially. Secondly, because it is determined by the higher-order dispersion, the phase-matching wavelength for RDW emission depends on the overall dispersion landscape of the waveguide, which in turn can be tuned by the gas fill---either the type of gas or the pressure. This is shown for some example parameters in \cref{fig:phasematching}(b). After the initial demonstration, this tuneability was quickly exploited to generate RDW pulses throughout the UV and visible~\cite{mak_tunable_2013} and into the far UV as low as \SI{110}{\nm}~\cite{ermolov_supercontinuum_2015}. The few-femtosecond pulse duration was also measured~\cite{ermolov_characterization_2016,brahms_direct_2019}, though not at the very shortest wavelengths; experimental confirmation of the duration of RDW pulses below \SI{200}{\nm} is still missing.

For far-UV RDW emission to become a practical light source, antiresonant fibres present two key challenges: firstly, the mechanism by which these fibres guide light naturally and inevitably creates wavelength regions in which light is in resonance with the cladding structure and not guided at all. This leads to gaps in the tuneability range, and strong dispersion at the edges of the resonances can disrupt the soliton self-compression process~\cite{tani_effect_2018}. Critically, the fibre resonances become more closely spaced at shorter wavelengths, so this issue is most significant for far-UV generation. Secondly, antiresonant fibres which guide light in the far-UV region are typically small, with a core radius of at most a few tens of \si{\um}. (Large-core antiresonant fibres typically only guide much longer wavelengths~\cite{fu_hundred-meter-scale_2022}, and it is not clear whether their advantages in propagation loss and bend resistance can be retained if they are modified to operate in the far UV.) The small core area limits the peak power and hence pulse energy which can be used to drive RDW emission. Importantly, the limiting factor is not the fibre itself, which can readily guide extremely intense laser pulses~\cite{lekosiotis_-target_2023}, but the gas inside it: because the need to phase-match RDW emission at some desired wavelength fixes the dispersion landscape, and hence the pressure for any given combination of gas species, driving wavelength and core size, the nonlinearity of the gas-filled waveguide is also fixed. Excessive peak power then leads to self-focusing and collapse of the beam inside the fibre~\cite{fibich_critical_2000} or to plasma generation through strong-field ionisation and breakup of the pulse~\cite{holzer_femtosecond_2011}. This limitation poses a challenge for ultrafast science applications, where the critical parameter is usually the single-shot pulse energy, but also for uses outside the laboratory, where average power is often more important. While far-UV RDW emission has been scaled to \SI{1}{\watt} average power in antiresonant fibres~\cite{kottig_generation_2017}, this required \si{\mega\hertz} pulse repetition rates. In this regime, the gas-filled fibre does not have sufficient time to fully relax between successive laser shots, so inter-pulse effects can disturb the dynamics~\cite{koehler_long-lived_2018,kottig_efficient_2020, sabbah_generation_2023} and reduce the conversion efficiency to the UV~\cite{kottig_generation_2017}. Nonlinear pulse compression and frequency shifting in antiresonant fibres have been demonstrated at even higher repetition rates~\cite{mak_compressing_2015,kottig_efficient_2020,tani_temporal_2022,schade_scaling_2021}, and it may be possible to extend this to far-UV generation, but this has yet to be demonstrated.

\section{HISOL: high-energy soliton dynamics in hollow capillary fibres}
\begin{figure*}
    \centering
    \includegraphics[width=17cm]{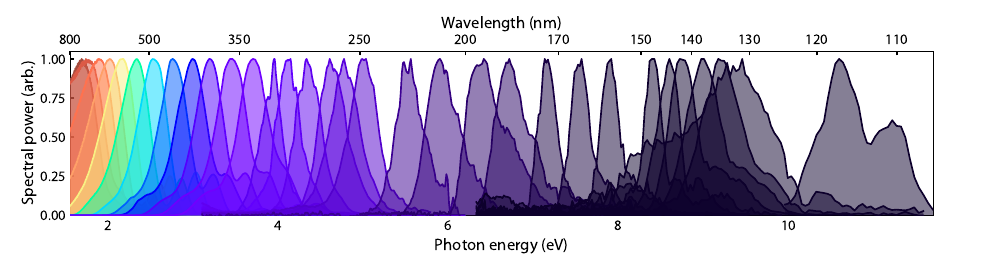}
    \caption{Wavelength-tuneable RDW emission across the near infrared, visible and ultraviolet. Data from refs.~\cite{travers_high-energy_2019, brahms_infrared_2020}.}
    \label{fig:RDW_tuning}
\end{figure*}
Initially, it was widely believed that antiresonant (which, at the time, meant kagom\'e-type~\cite{benabid_stimulated_2002}) fibres were the \emph{sine qua non} of observing soliton dynamics in gases. Hollow capillary fibres---essentially, simple glass tubes---were the obvious competing technology, and had been used for nonlinear pulse compression and other high-energy experiments for well over a decade by the time far-UV RDW emission was first demonstrated~\cite{nisoli_generation_1996}. However, antiresonant fibres offered a (supposedly) unique combination of significant anomalous dispersion and low-loss guidance across a huge bandwidth and this allowed for many important soliton dynamics to be observed~\cite{travers_optical_2024}. And indeed, at the core sizes of antiresonant fibres which guide far-UV light (typically \SIrange{20}{30}{\um} diameter), the propagation loss of even an ideal capillary is so high that any infrared pulse launched into it virtually disappears before it ever has the chance to self-compress. Antiresonant fibres do not present this problem. Somewhat embarrassingly, one of the present authors went so far as to claim that soliton dynamics were \emph{fundamentally impossible} to observe in capillaries~\cite{travers_ultrafast_2011,russell_hollow-core_2014}, only to (happily) prove himself wrong a few years later~\cite{travers_high-energy_2019}.

Although their loss properties are very different, antiresonant fibres and capillaries share one key characteristic: for most of the guidance window of an antiresonant fibre (and away from the resonances), the waveguide contribution to the dispersion can be extremely well approximated by the same model which describes that of capillaries essentially exactly~\cite{marcatili_hollow_1964, finger_accuracy_2014}. At large core sizes, the loss of capillaries becomes tolerable. So why were soliton dynamics not observed in capillary fibres immediately? The missing ingredient was a way of making the dispersion significant at the large core sizes for which capillaries can operate without huge losses (typically above \SI{100}{\um} diameter). For many years, capillaries were only used in circumstances in which the dispersion could be safely ignored, because they (mostly) focused on pulse compression in two stages---spectral broadening in the capillary followed by phase compensation with dispersive optics~\cite{nisoli_generation_1996}. This inherently implies the capillary being driven with \emph{long} pulses, which are less strongly affected by dispersion. In addition, capillaries have to be kept perfectly straight to avoid intolerable bend losses and mode mixing. With a traditional rigid capillary, which was the only available technology initially, the useful length was limited to around one metre (with rare and impressive exceptions~\cite{suda_generation_2005}). This had a double effect on the importance of dispersion: not only were traditional capillary systems too short for dispersion to have a significant effect, but this was compounded by the high gas pressure needed to provide sufficient nonlinearity for spectral broadening and pulse compression. The dispersion at the driving wavelength was thus most often very weakly anomalous or normal.

Three simple changes to a traditional hollow capillary fibre system bring dispersion into the picture and enable soliton dynamics~\cite{travers_high-energy_2019}: driving with short pulses, using a lower gas pressure, and increasing the fibre length. The longer fibre length, and hence by extension the lower gas pressure, is enabled by the transformative invention of stretched flexible hollow capillary fibre in 2008~\cite{nagy_flexible_2008}. Over several meters of propagation, even the weak dispersion of a gas-filled capillary with large core size (\SI{250}{\um} diameter) can have a significant effect on a 10-fs laser pulse, and this creates the conditions for soliton self-compression. For the purposes of this Perspective, the key result from the first experimental demonstration of this approach is the generation of \si{\uJ}-level few-femtosecond far-UV pulses between \SI{110}{\nm} and \SI{400}{\nm}~\cite{travers_high-energy_2019} (see \cref{fig:RDW_tuning}). Because of the dramatic overall energy scaling enabled by moving to larger cores, the dispersive wave contained nearly \SI{2}{\uJ} of energy even at the shortest wavelength, and this increased to over \SI{15}{\uJ} around \SI{220}{\nm}. The temporal profile of the UV pulses was not measured directly, but comparison with accurate numerical simulations suggested a pulse duration below \SI{2}{\fs}, and other work had previously confirmed similar durations in antiresonant fibres generated through the same mechanism~\cite{ermolov_characterization_2016,brahms_direct_2019}.

\begin{figure}
    \includegraphics[width=8.5cm]{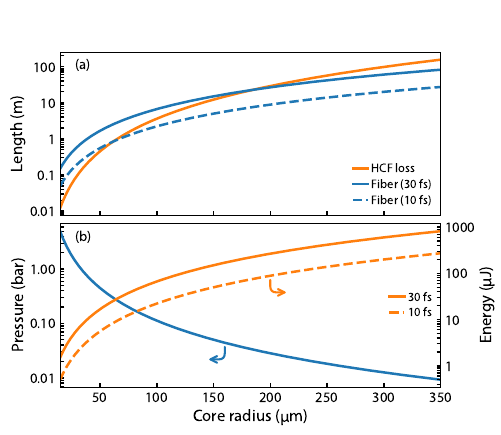}
    \caption{Scaling of soliton dynamics in gas-filled waveguides. (a) Fibre length scaled from the initial demonstration of far-UV RDW emission~\cite{joly_bright_2011} (\SI{15}{\um} core radius, \SI{5}{\bar} argon pressure, \SI{1.5}{\uJ} in \SI{30}{\fs} at \SI{800}{\nm}) when increasing the core radius (solid blue line) and when also including the effect of a shorter driving pulse (dashed blue line). The orange line shows the $1/\mathrm{e}$ loss length of a capillary fibre. When the fibre length is shorter than the loss length, soliton dynamics can be observed. (b) Scaled gas pressure (blue line, left axis) and pulse energy for the same two pulse durations (solid and dashed orange lines, right axis).}
    \label{fig:scaling}
\end{figure}
An alternative view of the origin of the HISOL concept, which is how the idea first came into being~\cite{travers_high-energy_2019,travers_optical_2024}, is as the consequence of scaling laws for soliton dynamics in gas-filled fibres. Starting with the original results in antiresonant fibres, observing similar dynamics in a larger core requires more energy, a longer fibre, and lower gas pressure, as shown in \cref{fig:scaling}. Crucially, the required fibre length increases more slowly than the propagation loss: the former scales with the square of the core radius, while latter scales inversely with its cube. The exact relations are derived explicitly for soliton dynamics in ref.~\cite{travers_high-energy_2019} and also include the effect of reducing the pulse duration, which dramatically reduces the required fibre length (dashed line in \cref{fig:scaling}(a)). However, the scaling rules for physical size, energy and pressure apply more generally to nonlinear optics in gases~\cite{heyl_scale-invariant_2016}. Nearly \emph{all} phenomena observed in antiresonant fibres could be scaled to capillaries, as the loss will always become tolerable at some point---the key question then becomes how much pulse energy and length is required to satisfy this condition. The only exceptions to this general rule are those dynamics which depend on the gas pressure directly---for instance, some Raman effects in molecular gases~\cite{bauerschmidt_dramatic_2015}---or on the specific resonance-induced dispersion profile of an antiresonant fibre~\cite{sollapur_resonance-enhanced_2017,tani_effect_2018,meng_controllable_2017}.

Although the first results already demonstrated many of the features which had long been sought after in a far-UV laser source---continuous tuneability, high pulse energy, and very short pulse duration---the hardest part of the work to finally fill the ultrafast far-UV technology gap was, and to some extent \emph{is}, still ahead. Over the past few years, our group has demonstrated the same process driven at much longer wavelengths, which allows for RDW emission from the far UV to the near infrared~\cite{brahms_infrared_2020} (see \cref{fig:RDW_tuning}); greatly reduced the space required for a capillary-based RDW source by driving with shorter pulses~\cite{brahms_high-energy_2019} and a more compact primary laser~\cite{brahms_efficient_2023}; extended far-UV RDW emission to circular polarisation~\cite{lekosiotis_ultrafast_2021} and higher-order capillary modes~\cite{brahms_soliton_2022}; investigated the effect of pressure gradients, which are required for direct delivery of RDW pulses to vacuum~\cite{brahms_resonant_2020}; and studied the noise properties of the RDW pulses~\cite{brahms_timing_2021}. This has taken us some way towards realising the potential of this technology, but significant challenges still remain; we will discuss these in detail later on in this Perspective.

\section{Applications of ultrafast far-ultraviolet sources}
New laser tools enable new applications, and we hope that filling the ultrafast far-UV technology gap will be particularly fruitful. From our vantage point, there are several key areas in which far-UV RDW sources can have immediate scientific and technological impact, and we will provide brief sketches of these below. Not all of these applications may prove worthwhile; perhaps none of them will. Our hope is that in any case, through the development and extensions motivated by these applications, RDW emission can mature into an easily accessible, flexible and reliable light-source technology for science and technology we cannot conceive of yet.

\subsection{Table-top ultrafast spectroscopy in excited neutral molecules}
The demands of next-generation experiments in ultrafast spectroscopy have been a key driver behind efforts to generate very short far-UV laser pulses~\cite{reiter_route_2010}. The goal is to study dynamics in photoexcited molecules at much faster timescales than traditional femtochemistry experiments. Direct measurements of electron motion in the first few femtoseconds after excitation promise to greatly improve our understanding of the primary processes which drive photochemistry, and may one day even open the door to controlling chemical properties and reactions on the ultrafast timescale~\cite{lepine_attosecond_2014,calegari_open_2023}. The simplest and most common experimental approach is pump-probe spectroscopy, in which a pump pulse excites the target molecule and a probe pulse measures the state of the system some time later. By repeating the experiment with different time delays between the two pulses, a picture of the dynamics can be constructed. Historically, photoelectron or -ion spectroscopy was the most common probe method in studies of molecular systems at the few-femtosecond scale~\cite{nisoli_attosecond_2017}. More recently, transient absorption spectroscopy, especially in the soft x-ray region, has become available as a powerful alternative~\cite{borrego-varillas_attosecond_2022,geneaux_transient_2019}. 

Because the time resolution of a pump-probe experiment is set by the duration of the two laser pulses, reaching the timescale of primary electronic processes in photochemistry requires advanced laser sources. The key technology for transient absorption spectroscopy at the few-femtosecond scale is extreme-ultraviolet (XUV) attosecond pulse generation by high-harmonic generation (HHG), which was recently recognised with the 2023 Nobel Prize in physics. XUV attosecond pulses can act as either the pump~\cite{calegari_ultrafast_2014} or the probe~\cite{cheng_reconstruction_2016} pulse. However, as XUV photon energies are sufficient to ionise any molecule, XUV-pump experiments are fundamentally unable to study dynamics in neutral systems. The most important ultrafast photochemical processes---for instance in vision, photosynthesis, and UV damage to DNA---happen in neutral systems, as ionising radiation from the sun does not penetrate our atmosphere. XUV or x-ray attosecond pulses could therefore be particularly useful when employed as the probe after resonantly exciting the target molecule at much lower photon energy. The technological challenge, then, is that the pump wavelength must match an absorption resonance in the target to enable single-photon excitation without ionisation. Virtually all molecules have strong resonances somewhere in the ultraviolet or visible. In particular, small polyatomic organic molecules, such as amino acids or nucleobases, exhibit valence electron transitions in the far UV.

Far-UV pump \emph{and} probe pulses have been used in photoelectron or -ion spectroscopy for many years~\cite{sorensen_femtosecond_2000,horio_full_2016,horio_probing_2009}. Here, resonant excitation is combined with photoionisation by another far-UV pulse. The charged particles (ions and/or electrons) created in this process can yield detailed information about intra-molecular energy conversion. For example, the photoelectron kinetic energy can directly encode the binding energy of the electronic state from which the electron was ionised; this can then be used to closely track motion on (and transitions between) potential-energy surfaces after photoexcitation. In such experiments, the pump and/or probe wavelengths are important not just to resonantly excite a system, but also to precisely address transitions to the appropriate cationic state with the probe. In contrast to UV-XUV photoelectron spectroscopy experiments~\cite{nugent-glandorf_ultrafast_2001}, the wavelengths can be chosen such that neither pulse causes single-photon ionisation, greatly reducing background signal in the measurement. Wavelength tuneability adds another important degree of freedom: the ultrafast dynamics started by the pump pulse may lead to multiple different end states, including different photoproducts, which cannot be readily distinguished with only a single probe wavelength. Scanning the probe wavelength across species-specific ionisation thresholds may relieve this ambiguity without the need for (much more complicated and time-consuming) photoelectron-photoion coincidence measurements. Finally, probing with a higher (but still far-UV) photon energy enables photoionisation from states reached only later during the photoexcited dynamics---thus ``extending the view along the reaction coordinate''~\cite{kotsina_improved_2021, townsend_mapping_2023}. Improving the time resolution will similarly allow for the investigation of faster and more complex dynamics, perhaps most importantly the passage through conical intersections, where the Born-Oppenheimer approximation breaks down and extremely fast internal energy conversion can occur. Here, charged-particle methods present an important drawback, in that the resolution in energy and time are intrinsically inversely linked. Faster experiments with extremely short, and hence broadband, far-UV pulses will therefore reveal important new insights which complement work with longer pulses but better spectral resolution~\cite{gierz_tracking_2015,kotsina_spectroscopic_2022}.

The need for very short (few-femtosecond) laser pulses in the far UV for both optical and charged-particle techniques has long been recognised as an important bottleneck in ultrafast molecular science~\cite{reiter_route_2010,lepine_attosecond_2014,calegari_open_2023,kotsina_improved_2021}. Far-UV laser pulses generated through RDW emission fit the requirements of cutting-edge ultrafast spectroscopy in three important respects: pulse duration, wavelength tuneability, and conversion efficiency. As mentioned already, few-femtosecond pulse duration is essential for the most advanced experiments. Experimentally measured far-UV RDW pulses have been around \SI{3}{\fs} long~\cite{ermolov_characterization_2016,brahms_direct_2019,reduzzi_direct_2023}. However, it is not yet clear whether this is a universal characteristic or due to the particularities of those experiments; different works vary considerably in their estimates of what is possible, with durations down to sub-\SI{2}{\fs} predicted from both simulations~\cite{travers_high-energy_2019,brahms_high-energy_2019} and experimentally measured bandwidths~\cite{brahms_efficient_2023,brahms_generation_2023}. Wavelength tuneability is similarly crucial to directly address relevant transitions wherever they may lie. Simply put, RDW emission is uniquely capable in this regard. By letting the sample dictate the wavelength, rather than the other way around, comparative studies and a much greater variety of samples will become feasible. The ability to reach wavelengths much shorter than the third harmonic of the primary driving laser is particularly important as ultrafast science moves towards ytterbium-based laser sources to increase the pulse repetition frequency and hence the speed of data acquisition~\cite{donaldson_breaking_2023}. Because these sources emit at \SI{1030}{\nm}, rather than \SI{800}{\nm} as delivered by long-established titanium-doped sapphire lasers, reaching the far UV without RDW emission requires cascaded nonlinear interactions~\cite{couch_ultrafast_2020}. Finally, RDW sources can be very efficient---we have demonstrated up to \SI{15}{\percent} conversion from the coupled driving pulse~\cite{travers_high-energy_2019}, and \SI{3.6}{\percent} total efficiency from the primary laser~\cite{brahms_efficient_2023}. This is more than simply ``nice to have'', especially for experiments with XUV or x-ray attosecond pulses. Because high-harmonic generation is famously inefficient, such attosecond light sources universally require most of the available driving pulse energy. If few-femtosecond far-UV pulses are to be combined with attosecond XUV sources in the same experiment, every photon counts.

\subsection{Ultrafast experiments with x-ray free-electron lasers}
X-ray free-electron lasers (XFELs) are by far the brightest sources of ultrafast short-wavelength radiation, with x-ray pulse energies many orders of magnitude higher than achievable with table-top systems~\cite{duris_tunable_2020,emma_first_2010}. The high photon flux they provide opens the door to far more photon-hungry experimental techniques, noteably single-shot x-ray scattering and imaging~\cite{odate_brighter_2022}. In contrast to spectroscopic approaches, these techniques can directly produce real-space images of ultrafast structural changes in molecules and solids~\cite{minitti_imaging_2015,johnson_ultrafast_2022}. The first two-colour pump-probe measurements purely with an XFEL source have already been demonstrated~\cite{barillot_correlation-driven_2021,guo_experimental_2024,li_attosecond-pump_2024}. However, because XFELs necessarily produce laser pulses at very high photon energy, they present the same limitation as table-top XUV pump-probe experiments. To excite neutral systems, XFELs must be coupled with secondary laser sources at much lower photon energy.

Although XFELs have now reached attosecond pulse duration~\cite{duris_tunable_2020,guo_experimental_2024}, the key challenge for most experiments is not the time resolution but the required flexibility and reliability. Infrared primary laser sources and traditional ultrafast frequency conversion techniques have been used extensively in XFEL experiments for this reason. Even moving to only slightly shorter wavelengths (e.g.~the fourth harmonic of a titanium-doped sapphire laser) can reveal new insights and resolve open questions~\cite{ruddock_deep_2019}. Tailoring the pump light to the target more precisely will open many more avenues in the same way---an area where the wide wavelength tuneability offered by RDW-emission sources could be uniquely powerful. However, XFELs are among the largest scientific facilities in existence and exceedingly expensive to run, so beam time is limited and very competitive. The best and most flexible ultrafast laser source is of no use if it cannot be relied upon to run stably through an entire experiment. While long-running experiments on the scale of a few hours have been carried out with many different ultrafast laser sources based on nonlinear optics in gas-filled HCFs~\cite{jager_tracking_2017,niedermayr_few-femtosecond_2022}, including HISOL sources~\cite{kotsina_spectroscopic_2022,brahms_decoupled_2024}, significant work remains in transforming them into reliable tools upon which such high-pressure experimental campaigns can be based routinely.

\subsection{Ultraviolet-driven strong-field physics}
The interaction between matter and very strong fields has been studied intensely ever since the advent of the laser enabled the tight confinement of electromagnetic energy in space and time. In addition to the insights gathered through strong-field processes themselves, arguably the most important contribution from this field of research is the development of attosecond light sources based on HHG. Although many early experiments in strong-field physics were carried out using excimer lasers in the far UV~\cite{mcpherson_studies_1987}, infrared-driven HHG and its many potential applications pushed the field firmly towards primary laser sources at longer wavelengths~\cite{ferray_multiple-harmonic_1988}. The use of infrared lasers for strong-field experiments became standard with the advent of amplified titanium-doped sapphire lasers (based on another Nobel-Prize-winning invention~\cite{strickland_compression_1985}), which operate at \SI{800}{\nm} and offer superior peak power and hence field strength. With even longer driving wavelengths, high harmonics at high photon energy in the x-ray region can be generated~\cite{shan_dramatic_2002} at the cost of a dramatic reduction in conversion efficiency~\cite{tate_scaling_2007,austin_strong-field_2012}.

Although infrared sources are still by far the most common, strong-field experiments with shorter wavelengths have turned up some unique results. A famous example is the ``ultraviolet surprise'': by driving HHG with an extremely intense ($\sim\!\SI{e16}{\watt\per\square\cm}$) femtosecond pulse at \SI{270}{\nm}, bright harmonics could be generated at much higher photon energies and with much better efficiency than would be expected~\cite{Popmintchev2015}. This result was attributed to an unusual phase-matching regime enabled by a balance between contributions from the plasma, ions, and neutral atoms in a multiply ionised gas. Reaching the necessary intensity in a loose-focusing geometry with a pulse duration around \SI{35}{\fs} required multi-millijoule pulse energy, which was made possible by a very high-energy titanium-doped sapphire amplifier and record-efficiency short-pulse third-harmonic generation~\cite{Popmintchev2015}. This regime is out of reach for most laser systems. While multi-colour control and/or enhancement schemes have been explored in detail~\cite{watanabe_two-color_1994,kim_highly_2005,brugnera_trajectory_2011} and some other experiments on purely UV-driven strong-field physics have been reported~\cite{mcpherson_studies_1987, wang_surprise_2023}, the technological barriers are formidable, and the ``ultraviolet surprise'' of highly efficient x-ray HHG has never been reproduced.

By generating very short far-UV laser pulses with high conversion efficiency, RDW emission is well placed to enable strong-field effects driven not only around \SI{270}{\nm} but at significantly shorter wavelengths. With few-femtosecond pulses and single-mode beam quality~\cite{travers_high-energy_2019,brahms_soliton_2022}, only a few hundred microjoules of pulse energy are required to significantly exceed the intensity used in the first ultraviolet surprise experiments. With the results we have obtained so far, we are still around one order of magnitude short: the brightest far-UV RDW pulses contain at most a few tens of microjoules. However, fundamental scaling laws tell us that HCF-based sources can be arbitrarily scaled up in energy---the only limit is the available pulse energy from the primary laser source and physical space in the laboratory~\cite{travers_high-energy_2019,heyl_scale-invariant_2016}. We are currently developing a scaled-up far-UV HISOL source (named XSOL) to investigate UV-driven strong-field physics in detail.

Focusing a few-femtosecond, few-hundred-microjoule far-UV dispersive wave to a diffraction-limited, few-hundred-nanometre focal spot will enter a regime far beyond that of gas-phase HHG experiments: even a conservative estimate predicts intensities over \SI{e19}{\watt\per\square\cm}. At this level, relativistic nonlinear optics experiments become feasible. One particularly interesting effect in this regime is relativistic harmonic generation from solid-density plasmas~\cite{gibbon_harmonic_1996}. This promises to create attosecond XUV and x-ray pulses with similar properties to gas-phase HHG, but potentially with orders of magnitude better conversion efficiency~\cite{heissler_multi-j_2015}. Although the first theoretical proposal already suggested the use of far-UV light to reach x-ray photon energies with a moderate harmonic order~\cite{gibbon_harmonic_1996}, available laser technology has restricted experimental demonstrations to infrared lasers~\cite{heissler_few-cycle_2012,heissler_multi-j_2015}. As with gas-phase HHG, driving relativistic harmonic generation with far-UV laser pulses is mostly unexplored, simply due to the lack of laser sources. XSOL will be able to access this regime. Importantly, the single-stage nonlinear frequency conversion should create pulses with the excellent pre-pulse contrast required for relativistic nonlinear optics---though the emphasis here is on \emph{should}, as this prediction has yet to be studied.

\subsection{Beyond ultrafast science}
Outside ultrafast photonics laboratories, one of the most important uses for laser radiation is material processing, from the industrial mass-production scale down to engineering research~\cite{choudhury_ultrafast_2014,ross_axi-stack_2024}. Far-UV lasers present several advantages in this field. Most immediately, shorter wavelengths can intrinsically be focused to smaller spots, improving the transverse resolution. Strong linear absorption by most materials can additionally confine the deposition of energy closer to the material surface, improving longitudinal resolution and reducing heat load~\cite{hodgson_ultrafast_2021}. Because the photon energy is high, far-UV light can also be used to machine materials which are transparent to longer wavelengths, such as glasses, through single-photon or few-photon absorption mechanisms~\cite{zhang_precise_1998,herman_laser_2000}. However, excimer lasers---which have been extremely successful in semiconductor photolithography, display manufacturing and some medical uses (e.g.~laser eye surgery)---cannot easily be used in this regime, because they generally do not produce high-quality beams which can be focused tightly. Recognising the potential, several major manufacturers now offer ultrafast UV and even far-UV laser sources, but these are still limited by their use of nonlinear crystals as the frequency conversion medium. Efficient far-UV generation techniques based on gas-filled hollow-core waveguides could potentially offer a step change in capability by generating near-perfect single-mode beams with much longer component lifetime. In terms of total output power, far-UV HISOL sources will most likely be outdone by FWM-based approaches, with which extraordinarily efficient conversion has been demonstrated~\cite{belli_highly_2019,belli_broadband_2020,forbes_efficient_2024}. However, the far-UV wavelength tuneability of dispersive waves could open an important additional degree of freedom. 

Far-ultraviolet light has found other important applications in the life sciences and biomedicine. For instance, although UV radiation is highly effective in combatting pathogens, even those which are resistant to antibiotics, it is also harmful to humans by causing DNA damage and cancer. However, this changes for very short wavelengths in the far UV, which can sterilise skin tissue and surgical sites without negative side effects~\cite{buonanno_207-nm_2013,buonanno_207-nm_2016}. Although excimer lamps can provide high-power incoherent light in this wavelength region, this does not allow for the efficient delivery of far-UV radiation in the same way as a spatially coherent laser beam. Here, the most important aspect is not the ultrafast pulse duration of far-UV laser sources based on nonlinear optics but the ability to reach this wavelength region at all. The same is true of applications in analytical science, for example in UV photodissociation for mass spectrometry of complex biomolecules~\cite{brodbelt_ultraviolet_2020}. Although the motivation for the HISOL approach was in significant part due to high-intensity or ultrafast applications, its other attributes may make it an important tool in many other areas.

\section{Current challenges}
Current experimental results already demonstrate that far-UV ultrafast laser sources based on RDW emission in HCF can make an impact in a variety of areas. However, several important challenges remain. What is the best way of delivering the short far UV pulse from inside the capillary to an experiment? How can we be sure that it remains intact through this process? How far can we push the pulse repetition rate and average power before running into the same issues encountered in antiresonant fibres~\cite{kottig_generation_2017,koehler_long-lived_2018}? Where are the ultimate limits of RDW emission? Below we will describe these challenges and current directions to overcome them.

\subsection{Spectral filtering}
Hollow-core fibres are ideal for far-UV RDW emission sources, because they provide the all-important anomalous dispersion contribution as well as extended interaction lengths. They also present one important drawback, however. To state the obvious, both the self-compressing soliton and the dispersive wave travel along the same waveguide. After the beam exits this waveguide, the far-UV pulse thus co-propagates with an intense and extremely broadband infrared and/or visible pulse. Even with efficient RDW emission, the peak power of this pulse can easily be several times greater than that of the RDW pulse. To avoid spurious signals in experiments, this must be removed and the dispersive wave isolated. Dichroic mirrors are the most common approach to filtering the driving pulse from nonlinear frequency conversion techniques, but current optical coating technology faces severe limitations when handling a combination of very large bandwidths and short wavelengths. While clean spectral filtering is possible even for sub-\SI{2}{\fs} pulses~\cite{galli_generation_2019}, the limited bandwidth prevents us from tuning such short pulses very far in wavelength. Nonetheless, for applications in which only one central wavelength is required, or where resources allow for the use of separate sets of optics, dichroic mirrors are by far the simplest solution.

A very broadband and nearly dispersion-free method of wavelength separation is offered by Brewster-angle reflection from mirrors made of silicon or similar materials~\cite{takahashi_high-throughput_2004}. Because the imaginary part of the refractive index of silicon increases very quickly for wavelengths below \SI{400}{\nm}, a silicon mirror at Brewster's angle for \SI{800}{\nm} ($\sim\!\ang{76}$) reflects between \SI{18}{\percent} and \SI{50}{\percent} of p-polarised light in the far UV but less than \SI{1}{\percent} in the visible. In addition to broadband dispersion-free filtering, the great advantage of this technique lies in its simplicity; high-quality silicon optics are widely available and cost-effective. However, several reflections are often required to fully remove the residual infrared and visible components~\cite{reiter_generation_2010}, which dramatically reduces the far-UV pulse energy. Here, the relatively high conversion efficiency and high energy scale of RDW emission in capillaries becomes a major advantage. A significant fraction of the driving pulse can be converted to the dispersive wave, so that less discrimination between the two spectral regions is required to efficiently isolate the far-UV pulse. Because the initial pulse energy is high, even inefficient methods leave sufficient photons for many experiments---but not \emph{all} experiments. If high-energy far-UV dispersive waves are to be used, for instance to drive nonlinear optics or for high-intensity applications in materials processing, more efficient, but still dispersion-free, methods will be required.


\subsection{Delivery of optimally short pulses}
A great variety of experimental and theoretical evidence suggests that far-UV dispersive-wave pulses can be very short indeed. It is not yet clear, however, which practical parameters determine their bandwidth, and hence the transform-limited duration. As just one example, generating RDW pulses between the far UV and the near infrared by driving at \SI{1800}{\nm}~\cite{brahms_infrared_2020}, we found that the transform-limited duration was essentially constant at \SI{3}{\fs} across the tuning range when using \SI{30}{\fs} driving pulses, but varied up to \SI{8}{\fs} when using pre-compressed \SI{16}{\fs} pulses. In many other frequency conversion schemes, the generated bandwidth is strongly influenced by the phase-matching bandwidth, which is determined by the rate of change of dephasing around the phase-matched frequency and the interaction length. Similar arguments can be made for RDW emission~\cite{brahms_soliton_2022,kottig_mid-infrared_2017}, but two important caveats prevent accurate predictions: firstly, we already know that the phase-matching condition for RDW emission is only approximate, and secondly, the interaction length is unclear due to the dynamical nature of the process. Because the RDW characteristics are only indirectly coupled to the driving pulse, it is currently hard to predict the RDW duration without detailed numerical simulations. A deeper theoretical understanding of the dynamics could help in the design of RDW emission systems to produce the shortest possible far-UV pulses.

The shortest pulse in the world is useless if it cannot be delivered to an experiment. Although RDW pulses can be extremely short \emph{at the point of generation} inside the capillary, this is no guarantee that they even make it out of the fibre intact. Quite the opposite, in fact: phase-matching of RDW emission relies on the frequency of the dispersive wave lying in the opposite dispersion regime to the driving pulse~\cite{erkintalo_cascaded_2012}. Because the dispersion of all materials and even gases increases for shorter wavelengths in the far UV, the effect can be significant even in an otherwise weakly dispersive system like a capillary. If left to propagate in the gas-filled waveguide, an RDW pulse can stretch to several times its initial duration~\cite{brahms_resonant_2020}. Transmission through any optical windows or even short beam paths in atmospheric air have even more disastrous effects, creating picosecond pulses out of few-femtosecond ones~\cite{ermolov_characterization_2016,travers_high-energy_2019,brahms_infrared_2020}. As with dichroic mirrors, dispersion-compensating optics such as chirped mirrors are a major challenge for current optical technology, and are only available for a select few wavelength bands (centred mostly on the third harmonic of titanium-doped sapphire lasers) and without the bandwidth required for few-femtosecond pulses. (At slightly longer wavelengths, dispersive optics have already been used to deliver very short RDW pulses to a characterisation setup~\cite{zhang_measurements_2022}.) The only way of maintaining short pulses is thus to avoid dispersion altogether.

The key technique to achieve dispersion-free pulse delivery is a negative (decreasing) pressure gradient along the capillary, which allows for subsequent in-vacuum beam transport without any windows. Filling gas into the entrance of the waveguide while evacuating its exit leads to a characteristic square-root dependence of the pressure along the length~\cite{suda_generation_2005}, and good vacuum can be maintained on the output side~\cite{ermolov_low_2013}. The gradient is inverted with respect to the more common approach for nonlinear pulse compression, in which the \emph{entrance} is evacuated to avoid stability or efficiency issues resulting from nonlinear effects on the beam as it is coupled into the capillary~\cite{suda_generation_2005,Robinson2006}. Based on that work, one may wonder whether the \emph{decreasing} gradient causes severe issues due to exactly those nonlinear effects which an \emph{increasing} gradient is designed to avoid. However, it is important to note that the nonlinearity in a gas-filled capillary for soliton self-compression is almost always much lower than in a pulse compressor, because the gas pressure is relatively low to keep the dispersion anomalous. For example, in the experimental system which was used for the first demonstration of the HISOL approach~\cite{travers_high-energy_2019}, nonlinear degradation of the coupling efficiency only occurs at pressures well beyond the soliton regime~\cite{grigorova_dispersion_2023}.

Using a decreasing gradient, compressed, few-femtosecond RDW pulses in the 200--300~nm region have already been delivered to pulse-characterisation experiments~\cite{brahms_direct_2019,reduzzi_direct_2023}. However, for even shorter wavelengths, even shorter pulses, and---most importantly---more complicated experiments, the optics used to deliver the beam to its destination become important. Very broadband laser pulses are most commonly transported with metallic (most often silver) mirrors with a protective silica layer, which exhibit essentially negligible dispersion in the visible and infrared. In the far UV, silver must be replaced by aluminium, but this is not enough for few-femtosecond pulses: a typical protective silica layer is over \SI{100}{\nm} thick~\cite{jobst_optical_2014}, which leads to a group-velocity dispersion of \SIrange{1}{5}{\fs\squared} per reflection. Comparing this to the square of the pulse duration---around \SIrange{4}{10}{\fs\squared}---it is clear that significant pulse stretching occurs with only a few reflections. Aluminium mirrors with magnesium-fluoride protective layers must therefore be used. These are primarily designed to reflect very short wavelengths, down to \SI{120}{\nm}, but the dispersion is reduced by around an order of magnitude as a welcome side effect.  

\subsection{Few-femtosecond, far-UV pulse characterisation}
\begin{figure}
    \centering
    \includegraphics[width=8.5cm]{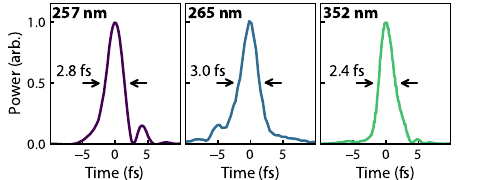}
    \caption{Far-ultraviolet dispersive-wave pulses measured with in-vacuum SD FROG by Reduzzi \emph{et al.}~\cite{reduzzi_direct_2023}.}
    \label{fig:reduzzi}
\end{figure}
Given how fragile they are, all ultrafast laser pulses are uncompressed until proven otherwise. This is especially important for RDW pulses: strong material dispersion, limitations of mirror technology, and the generation mechanism itself can all destroy a few-femtosecond pulse very easily. Pulse characterisation is essential to determine the time resolution of ultrafast experiments and to calculate other important metrics, such as the peak intensity. Measuring few-femtosecond pulses at any wavelength is a significant technical challenge, as is measuring pulses of any duration at short wavelengths. Combining both makes for a very hard metrology problem. Far-UV RDW pulses have been characterised with three different all-optical techniques: transient grating cross-correlation frequency-resolved optical gating (TG-XFROG)~\cite{ermolov_characterization_2016}, difference-frequency-generation time-domain ptychography (DFG-TDP)~\cite{brahms_direct_2019}, and self-diffraction (SD) FROG~\cite{reduzzi_direct_2023} (see \cref{fig:reduzzi}). Each of these presents its own advantages and drawbacks: SD FROG is experimentally simple, but requires very thin samples and provides limited bandwidth; DFG-TDP uniquely produces a signal at a different wavelength, increasing signal levels, but is also limited in bandwidth by phase-matching constraints; TG-XFROG is the most broadband but requires a complex optical arrangement. The use of a transmissive medium (a nonlinear crystal or a glass sample) furthermore adds uncertainty and limitations to such all-optical approaches. Even if it is only \SI{10}{\um} thick, a silica sample adds around \SI{4}{\fs\squared} at \SI{200}{\nm}, so the RDW pulse changes while it is being measured. In addition, silica absorbs below \SI{160}{\nm}, so other materials like magnesium fluoride must be used, but these are not widely available in such thin samples. One promising avenue for completely dispersion-free pulse characterisation would be to exploit nonlinear effects at surfaces, which have already been used in other spectral ranges~\cite{lin_spectral_2004,anderson_gold-spider:_2008}. However, it is not clear whether this approach is easily transferred to the far UV, and whether sufficient signal could be obtained in an experiment.

As with the generation mechanism itself, solid media may have to be abandoned in favour of gases to fully characterise the shortest pulses at the shortest far-UV wavelengths. Autocorrelation measurements based on multi-photon ionisation have been used to measure pulses as short as \SI{3}{\fs}~\cite{graf_intense_2008,reiter_generation_2010}. In this technique, two copies of the far-UV pulses (created with a split mirror) are overlapped in a gas target and the total photoelectron or -ion yield is recorded as a function of delay between them. Alternatively, cross-correlation measurements with a short infrared pulse have measured both far-UV dispersive waves~\cite{kotsina_spectroscopic_2022} and sub-\SI{2}{\fs} pulses created by third-harmonic generation~\cite{galli_generation_2019}. The obvious drawback of such spectrally integrated measurements is that they do not fully characterise the pulse, only its duration. For many applications this may be sufficient, and is therefore likely to remain the simplest and most flexible gas-based characterisation method. If, however, the pulse is not as short as expected, or we need to know the exact pulse shape, or we want to investigate the generation dynamics themselves, we must combine a dispersion-free gas-based method with the ability to measure the phase of the pulse. One possible approach is to use tunnelling ionisation with a perturbation for the time-domain observation of an electric field (TIPTOE)~\cite{park_direct_2018}. Here, ionisation driven by a strong replica of the unknown pulse is perturbed by interference with a weak replica, and the total electron yield as a function of the delay between the two pulses encodes the full electric field. TIPTOE has already been used to measure single-cycle, 3-fs pulses in the infrared~\cite{tsai_nonlinear_2022} and can likely be used at even shorter duration. The key question is whether it can be extended to short wavelengths. At far-UV photon energies, multi-photon ionisation begins to dominate over tunnelling ionisation, so the basic working mechanism of TIPTOE may be disabled. 

\subsection{Repetition-rate scaling}
Whether in scientific or industrial applications, higher average power and pulse repetition rate increase processing speeds---more photons are almost always useful. In far-UV-pumped ultrafast spectroscopy experiments, a high pulse repetition rate may not only be useful, it may be \emph{required} to overcome signal-to-noise issues created by the use of intense short-wavelength pulses. Only two or three far-UV photons are required to ionise most molecules. When the goal is to study dynamics in neutral systems, the pump intensity is therefore limited. Combined with a very short pulse duration, this in turn limits the pulse energy, and hence the excitation fraction in the experimental sample. As the differential absorption (pumped vs.~un-pumped) in a pump-probe experiment is directly determined by this fraction, very low-noise measurements may be required to record any meaningful signal at all. Besides a stable source of probe pulses, the main way of acquiring the necessary statistics is to increase the pulse repetition rate.

Ytterbium-based ultrafast lasers are now a mature technology and can deliver orders of magnitude more average power than traditional titanium-doped sapphire lasers at repetition rates of hundreds of \si{\kilo\hertz} to \si{\mega\hertz}~\cite{muller_35_2018,grebing_kilowatt-average-power_2020}. Whether or not the full repetition rate and average power delivered by such sources can be used to drive far-UV RDW emission is not yet clear, however. In antiresonant fibres, long-lived changes in the transverse refractive index profile caused by strong-field ionisation during soliton self-compression were found to limit the average power and repetition rate of far-UV RDW emission~\cite{koehler_long-lived_2018,kottig_generation_2017}. The details of the \si{\us}-scale dynamics following the laser pulse are complex and involve both the gas temperature and density as well as vibrations induced in the cylindrical resonators of single-ring antiresonant fibres~\cite{koehler_long-lived_2018}. In the absence of such resonators, capillaries may be able to handle higher pulse repetition rate than antiresonant fibres, but experimental evidence is still lacking. The highest repetition rate achieved so far is \SI{50}{\kilo\hertz}~\cite{brahms_efficient_2023}. However, because the single-pulse energy is significantly higher, this source already provided around one third of the far-UV average power of the highest-power result in antiresonant fibres~\cite{kottig_generation_2017}. Capillaries have been used to compress laser pulses at hundreds of watts of average power and \SI{100}{\kilo\hertz} repetition rate~\cite{nagy_generation_2019}, and we expect that similar power levels should be achievable with soliton dynamics. Such a source would provide entirely unique laser radiation even without considering the ultrafast pulse duration: tens of watts of average power in the far UV with full spatial coherence, near-perfect beam properties and wide-ranging wavelength tuneability.

\subsection{Short-wavelength limits of RDW emission}
RDW emission sources have already shown themselves to be uniquely capable of generating very short pulses at very short wavelengths down to \SI{110}{\nm}. The natural question is: how low can we go? At time of writing, it is not clear what determines the short-wavelength limit of RDW emission. It stands to reason that regardless of all other aspects, the transparency of the lightest noble gas---helium---is a fundamental barrier. However, this would place the limit around \SI{50}{\nm}, more than an octave higher than what has been achieved so far.

The clearest hints to the practical limitations come from experiments at different wavelengths: RDW emission at \SI{110}{\nm} has only been obtained by driving at \SI{800}{\nm}~\cite{ermolov_supercontinuum_2015,travers_high-energy_2019}. With a \SI{1030}{\nm} driving pulse, RDWs as low as \SI{145}{\nm} have been observed~\cite{brahms_generation_2023}, though that experiment was limited by the detector. The shortest published RDW spectrum generated with even longer driving wavelengths is centred at \SI{210}{\nm}~\cite{brahms_infrared_2020}. In unpublished experiments, we have pushed RDW emission driven at \SI{1800}{\nm} just below \SI{190}{\nm} but were unable to go any further; all attempts at tuning to even shorter wavelengths resulted in the loss of all UV generation. It appears likely that the driving wavelength plays a key role: because the spectrum of the self-compressing soliton must extend to the phase-matched wavelength before RDW emission can occur, generating frequencies further away from the driving field requires more extreme soliton dynamics. As the RDW is tuned to shorter wavelengths, higher-order nonlinear effects---most importantly strong-field ionisation and the resulting soliton self-frequency blue-shift~\cite{holzer_femtosecond_2011}---become more prominent and may prevent RDW emission from occurring at all. However, the detail of this mechanism, whether factors other than the driving wavelength are important, and whether this barrier can be overcome, is yet to be understood.

\section{Concluding remarks}
The need for flexible ultrafast laser sources in the far UV has motivated major experimental efforts for more than two decades. RDW emission from high-energy solitons promises to be a turning point in this development, simultaneously satisfying many of the most important requirements for cutting-edge ultrafast science. On a technical level, HISOL represents a relatively small advance based on a mountain of previous work. Most of the nonlinear optical dynamics at play are precisely the same as those observed at lower energies in solid-core and antiresonant hollow-core waveguides, and at first sight, a HISOL source does not look materially different to many other ultrafast optics experiments employing hollow capillary fibres. However, the capabilities gained by scaling up low-energy far-UV sources and removing the limitations of antiresonant waveguides are significant. One of the present authors, when attempting to integrate a far-UV source based on RDW emission in an antiresonant fibre into a collaborator's experiment, was once told to ``come back when you have a microjoule''. That has now been achieved---and then some more on top.

High-energy few-femtosecond far-UV dispersive waves from capillaries fill an important technology gap in ultrafast science and will additionally enable high-power far-UV laser sources using only moderate pulse repetition rates. However, there are and (we expect) always will be applications where small-core antiresonant fibres reign supreme. The most obvious is the opposite scaling direction to the HISOL approach: soliton self-compression and RDW emission driven by very \emph{low}-energy laser pulses~\cite{xiong_low-energy-threshold_2021, sabbah_ultra-low_2024}. Pushing the driving-pulse energy down instead of up, to well below \SI{1}{\uJ}, may one day allow for direct far-UV generation from ultrafast laser oscillators~\cite{liu_megawatt_2017,sidorenko_generation_2020,chen_efficient_2022} which would open the door to the creation of far-UV frequency combs. With carefully designed (and potentially post-processed) antiresonant fibres, core diameters of \SI{10}{\um} or even less and sub-100-nJ infrared pulse energies can be used~\cite{xiong_low-energy-threshold_2021,sabbah_ultra-low_2024}. In this regime, capillaries are simply hopelessly lossy, and even pre-compressed driving pulses cannot remedy this.

The key difference between RDW emission and other frequency conversion techniques lies in the subtle and complex evolution of the infrared driving pulse before it generates the dispersive wave, which uncouples the far-UV pulse properties from those of the primary laser source and allows for extremely short pulses to be created from much longer ones. However, while this makes RDW sources uniquely flexible, it also creates significant challenges in the practical implementation: because the self-compression process must be optimised in order for RDW emission to produce the desired result, choosing the correct experimental parameters can be very complicated. For example, compared to pulse compressors based on gas-filled HCF, the importance of the dispersion in addition to nonlinearity places significant constraints on the range of gas pressures and core sizes which can be used. Even with a growing body of literature exploring the details of HISOL sources, a barrier to widespread adoption by any research group who could benefit from them remains. As a first step to facilitate access to the technology where required, we have published our numerical simulation code, which has been extensively benchmarked against experimental results and can make quantitative predictions about nonlinear optical dynamics in gas-filled HCF and other sytems without any free parameters~\cite{brahms_lunajl_2021}. In the future we will additionally provide comprehensive design guidelines for HISOL sources. In this Perspective, we have laid out where HISOL sources came from, where they are now, and where we think they can go in the future. This approach will make new enabling tools available for cutting-edge ultrafast science and many other areas.

\section*{Acknowledgements}
We would like to thank all members of the Laboratory of Ultrafast Physics and Optics at Heriot-Watt University as well as our numerous collaborators for their dedication and hard work which made the work reviewed here possible. We would also like to thank Dave Townsend for fruitful discussions regarding this Perspective.

This work was funded by the European Research Council (ERC) under the European Union's Horizon 2020 research and innovation programme: Starting Grant agreement HISOL no.~679649 and ERC Consolidator Grant XSOL no.~101001534 and by the United Kingdom's Engineering and Physical Sciences Research Council: Grant agreement EP/T020903/1. CB acknowledges support from the Royal Academy of Engineering through Research Fellowship No. RF/202122/21/133. JCT acknowledges support by the Institution of Engineering and Technology (IET) through the IET A F Harvey Engineering Research Prize.

\section*{Data availability}
Data sharing is not applicable to this article as no new data were created or analyzed in this study.

\section*{Author Declarations}
The authors have no conflicts to disclose.

\section*{References}
\bibliography{bibliography}
\end{document}